\documentclass[conference]{IEEEtran}
\usepackage{amsmath,graphicx,times}

\IEEEoverridecommandlockouts    % using \thanks

\begin{document}

\title{\ \\ \LARGE\bf Variable-Length Haplotype Construction for Gene-Gene Interaction Studies \thanks{A. Assawamakin, N. Chaiyaratana and C. Limwongse are with the Division of Molecular Genetics, Department of Research and Development, Faculty of Medicine Siriraj Hospital, Mahidol University, 2 Prannok Road, Bangkoknoi, Bangkok 10700, Thailand.} \thanks{N. Chaiyaratana, S. Sinsomros and P. Youngkong are with the Department of Electrical Engineering, Faculty of Engineering, King Mongkut's University of Technology North Bangkok, 1518 Piboolsongkram Road, Bangsue, Bangkok 10800, Thailand (phone: 66-2913-2500 ext 8410; fax: 66-2585-6149; email: nchl@kmutnb.ac.th).} \thanks{P.-T. Yenchitsomanus is with the Division of Medical Molecular Biology, Department of Research and Development, Faculty of Medicine Siriraj Hospital, Mahidol University, 2 Prannok Road, Bangkoknoi, Bangkok 10700, Thailand.} \thanks{This article has been accepted for publication in {\bf IEEE Engineering in Medicine and Biology Magazine} on 19 November 2007.}
}

\author{A. Assawamakin, N. Chaiyaratana, {\it Member, IEEE}, C. Limwongse, S. Sinsomros, P.-T. Yenchitsomanus \\ and P. Youngkong}
\maketitle

\begin{abstract}
This paper presents a non-parametric classification technique for identifying a candidate bi-allelic genetic marker set that best describes disease susceptibility in gene-gene interaction studies. The developed technique functions by creating a mapping between inferred haplotypes and case/control status. The technique cycles through all possible marker combination models generated from the available marker set where the best interaction model is determined from prediction accuracy and two auxiliary criteria including low-to-high order haplotype propagation capability and model parsimony. Since variable-length haplotypes are created during the best model identification, the developed technique is referred to as a variable-length haplotype construction for gene-gene interaction (VarHAP) technique. VarHAP has been benchmarked against a multifactor dimensionality reduction (MDR) program and a haplotype interaction technique embedded in a FAMHAP program in various two-locus interaction problems. The results reveal that VarHAP is suitable for all interaction situations with the presence of weak and strong linkage disequilibrium among genetic markers.

\end{abstract}
%
% 1. Introduction
%
\section{Introduction}
Genetic epidemiology is a research field which aims to identify genetic polymorphisms that involve in disease susceptibility. Usual candidate polymorphisms include restriction fragment length polymorphisms (RFLPs), variable numbers of tandem repeats (VNTRs) and single nucleotide polymorphisms (SNPs). In recent years, SNPs are the most common choices due to simplicity and cost reduction in identification protocols. SNPs in diploid organisms are excellent bi-allelic genetic markers for various studies including genetic association, gene-gene interaction and gene-environment interaction. The availability of multiple SNPs on the same gene can also lead to haplotype analysis where genotypes of interest can be phased into pairs of haplotypes.

Traditional techniques for identification of relationship between a single SNP and disease susceptibility status involve various univariate statistical tests including $\chi^2$ and odds ratio tests~\cite{Lewis,Montana}. However, many complementary computational techniques have been developed in the past decade to handle problems that involve multiple SNPs. Heidema et al.~\cite{Heidema} have categorised these multi-locus techniques, which are capable of identifying a candidate SNP set from possible SNPs, into parametric and non-parametric methods. Examples of parametric method cover logistic regression techniques~\cite{Nagelkerke} and neural networks~\cite{Ritchie}. On the other hand, examples of non-parametric method include a set association approach~\cite{Hoh}, combinatorial techniques~\cite{Nelson,Culverhouse,Hahn} and recursive partitioning techniques~\cite{Lunetta,Bureau}. In some of mentioned parametric~\cite{Nagelkerke,Ritchie} and non-parametric~\cite{Hahn} methods, pattern recognition and classification approaches have been successfully implemented as their core engines.

In addition to single and multiple SNP analysis, haplotype analysis has also gained attention from genetic epidemiologists. Haplotypes provide a record of evolutionary history more accurately than individual SNPs. Further, haplotypes can capture the patterns of linkage disequilibrium (LD)---a phenomenon where SNPs that are located in close proximity tend to travel together---in genome more accurately. Therefore, haplotypes may enable susceptibility gene identification in complex diseases more effectively than individual SNPs~\cite{Silverman}. In lieu of this evidence, haplotype analysis should also be considered in addition to direct genotype analysis. Many computational techniques use haplotypes, which are inferred from multiple SNPs, as problem inputs. For instance, Sham et al.~\cite{Sham} proposes a logistic regression technique that produces a mapping model between haplotypes and disease status while Becker et al.~\cite{Becker} combine haplotype explanation probabilities of given genotypes from multiple gene or unlinked region data into a scalar statistic for a univariate test. Nonetheless, haplotypes have rarely been used as inputs for non-parametric classifiers for genetic association and interaction studies.

In this paper, a variable-length haplotype construction for gene-gene interaction (VarHAP) technique is proposed. The technique will involve non-parametric classification where haplotypes inferred from multiple SNP data are the classifier inputs. The chosen architecture for non-parametric classifier is the multifactor dimensionality reduction (MDR) technique~\cite{Hahn}. Similar to the original MDR technique, the proposed technique would be able to identify appropriate candidate SNPs from possible SNPs and can be used in case-control genetic interaction studies. However, the technique would also be able to handle the situation where disease susceptibility is detectable in different haplotype backgrounds.

The organisation of this paper is as follows. In section~\ref{sec:MDRandHaplotype}, a brief explanation of MDR and the techniques for inferring haplotypes and obtaining haplotype explanation probability is given. The proposed VarHAP technique is then described in section~\ref{sec:VarHAP}. The test data and their description is given in section~\ref{sec:DataSets}. Next, the results and discussions are described in section~\ref{sec:Results}. Finally, the conclusions are drawn in section~\ref{sec:Conclusions}.
%
% 2. MDR Technique, Haplotype Inference and Haplotype Explanation Probability
%
\section{MDR, Haplotype Inference and Haplotype Explanation Probability}
\label{sec:MDRandHaplotype}
%
% 2.1. MDR
%
\subsection{MDR}
\label{subsec:MDR}
\begin{figure}[t!]
\centering
\includegraphics[width=5.0cm]{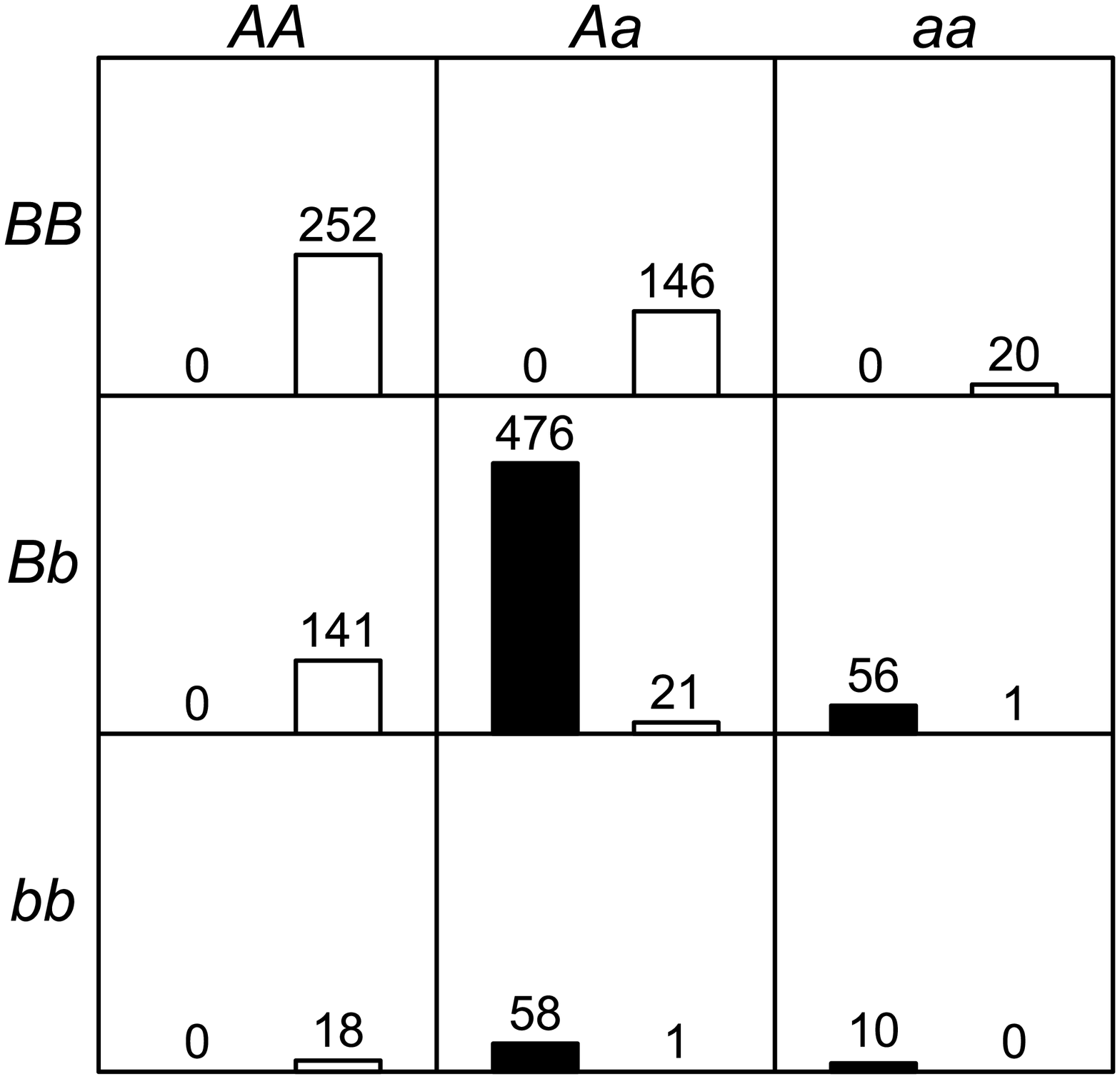}
\caption{An MDR decision table which is constructed using 1,200 case-control samples. The genotype of each sample is determined from two SNPs. The table consists of nine cells where each cell represents a unique genotype. The left (black) bar in each cell represents the number of case samples while the right (white) bar represents the number of control samples. The cells with genotypes {\it AaBb}, {\it aaBb}, {\it Aabb} and {\it aabb} are labelled as predisposing genotypes while the cells with genotypes {\it AABB}, {\it AaBB}, {\it aaBB}, {\it AABb} and {\it AAbb} are labelled as protective genotypes.}
\label{fig:MDRTable}
\end{figure}

MDR is a classifier-based technique that is capable of identifying the best genetic marker combination among possible markers for the separation between case and control samples. Similar to other classification systems, a $k$-fold cross-validation technique provides a means to determine the classification accuracy of the candidate marker model. Basically, the combined case and control samples are randomly divided into $k$ folds where $k-1$ folds of samples are used to construct a decision table for the classifier while the remaining fold of samples is used to identify the prediction capability of the constructed decision table. The decision table construction and testing procedure is repeated $k$ times. Hence, the samples in each fold will always be utilised both to construct and to test the decision table. The number of cells in a decision table is given by $G^{n_c}$ where $n_c$ is the number of candidate markers selected from possible markers and $G$ is the number of possible genotypes according to the marker. For a SNP, which is a bi-allelic marker, $G$ is equal to three. During the decision table construction, each cell in the table is filled with case and control samples that have their genotype corresponds to the cell label. The ratio between numbers of case and control samples will provide the decision for each cell whether the corresponding genotype is a disease-predisposing or protective genotype. An example of decision table construction is illustrated in Figure~\ref{fig:MDRTable}. The prediction accuracy of the decision table is subsequently evaluated by counting the numbers of case and control samples in the testing fold that their disease status can be correctly identified using the constructed decision rules. The process of decision table construction and evaluation must be cycled through all or some of possible $2^{n_m}-1$ combinations where $n_m$ is the total number of available markers in the study. The best genetic marker combination is determined from three criteria: prediction accuracy, cross-validation consistency and a sign test {\it p}-value. Each time that a testing fold is used for prediction accuracy determination, the accuracy of the interested marker combination model can be compared with that from other models that also contain the same number of markers. The model that consistently ranks the first in comparison to other choices with the same amount of markers would have high cross-validation consistency. The non-parametric sign test {\it p}-value is calculated from the number of testing folds with accuracy greater than or equal to 50\%. This single-tailed {\it p}-value is given by
%
% Equation 1 Sign test p-value
%
\begin{equation}
p = \sum_{i=n_a}^{n_f}\binom{n_f}{i}\left(\frac{1}{2}\right)^{n_f}
\label{eq:SignTest}
\end{equation}
where $n_f$ is the total number of cross-validation folds and $n_a$ is the number of cross-validation folds with testing accuracy $\geq 50\%$~\cite{Hahn}. Among three criteria, prediction accuracy is the main criterion for decision making while the other criteria are only used as auxiliary measures. Cross-validation consistency generally confirms that the high rank model can be consistently identified regardless of how the samples are divided for cross-validation. On the other hand, a sign test {\it p}-value indicates the number of testing folds with acceptable prediction accuracy and hence describes the usability of the model in the classification task. In the situation where two or more models with different number of markers are equally good in terms of prediction accuracy, cross-validation consistency and sign test {\it p}-value, the most parsimonious model---the combination with the least number of markers---will be the best model.
%
% Fig. 1. MDR table
%

%
% 2.2. Haplotype Inference
%
\subsection{Haplotype Inference}
\label{subsec:HaplotypeInfer}
With the availability of multiple SNPs from the same gene, haplotypes can be inferred from given genotypes. Let `\texttt{0}' and `\texttt{1}' denote the major (common) and minor (rare) alleles at a SNP location in a haplotype. A genotype can then be represented by a string, which consists of three characters: `\texttt{0}', `\texttt{1}' and `\texttt{2}'. In the genotype string, `\texttt{0}' denotes a homozygous wide-type site, `\texttt{1}' denotes a heterozygous site and `\texttt{2}' denotes a homozygous variant or homozygous mutant site. A genotype with all homozygous sites or single heterozygous site can always be phased into one pair of haplotypes. On the other hand, a genotype with multiple heterozygous sites can be phased into multiple haplotype pairs. For example, genotype \texttt{0102} leads to haplotypes \texttt{0001} and \texttt{0101} while genotype \texttt{0112} leads to two possible haplotype pairs: \texttt{0001/0111} and \texttt{0011/0101}. Many algorithms exist for haplotype inference~\cite{Excoffier,Stephens,Niu}. In this paper, an expectation-maximisation algorithm~\cite{Excoffier} is the chosen technique due to its simplicity and implementation efficacy. Regardless of the inference technique employed, the usual result from an inference algorithm covers haplotype frequencies and possible haplotype phases of each genotype.
%
% 2.3. Haplotype Explanation Probability
%
\subsection{Haplotype Explanation Probability}
\label{subsec:HaplotypeExplain}
In a genomic region with multiple heterozygous sites, multiple pairs of haplotypes can be inferred from a given genotype. The probability of a genotype to be phased into one specific pair of haplotypes would depend on the frequencies of haplotypes constituting the pairs~\cite{Becker}. This probability is given by
%
% Equation 2 Haplotype probability
%
\begin{equation}
w_{ij} = \frac{f_if_j}{\sum_{(h_k,h_l)\in H}f_kf_l}
\label{eq:Probability}
\end{equation}
where $w_{ij}$ is the probability for haplotype pair $ij$, $f_i$ denotes the frequency of the $i$th haplotype, $h_k$ is the $k$th haplotype and $H$ represents the set of haplotype explanations which are compatible with the genotype of interest. For example, genotype \texttt{0110} can be phased into two haplotype pairs: \texttt{0000/0110} $(h_1/h_4)$ and \texttt{0010/0100} $(h_2/h_3)$. If the frequencies for haplotypes \texttt{0000}, \texttt{0010}, \texttt{0100} and \texttt{0110} are respectively 0.5, 0.2, 0.2 and 0.1, the probabilities for the pairs \texttt{0000/0110} and \texttt{0010/0100} are 0.556 and 0.444. Obviously, the probability of a genotype with all homozygous sites or single heterozygous site to be phased into a pair of haplotypes would be equal to one. In genetic interaction studies where the number of genes or unlinked regions is greater than one, the haplotype explanation probabilities from all regions can be combined together. An overall contribution by one sample to haplotype configuration $(h_j^1,h_j^2,\dots,h_j^{n_u})$ in a study with   $n_u$ genes/unlinked regions is given by
%
% Equation 3 Haplotype Contribution
%
\begin{equation}
c_{(h_j^1,h_j^2,\dots,h_j^{n_u})} = 2\prod_{i=1}^{n_u}w_{jk}^i\frac{(1+\delta_{jk}^i)}{2}
\label{eq:Contribution}
\end{equation}
where $c_{(h_j^1,h_j^2,\dots,h_j^{n_u})}$ is the contribution value and $\delta$ is defined as $\delta_{jk} = 1$ for $j = k$ and $\delta_{jk} = 0$ for $j \neq k$. In the previous example where haplotypes from only one region are considered, $c_{h_1} = 0.556$  , $c_{h_2} = 0.444$, $c_{h_3} = 0.444$ and $c_{h_4} = 0.556$. Notice that the sum of contribution values is equal to two; this reflects the fact that each genotype is made up from two haplotypes. Becker et al.~\cite{Becker} use this contribution value in the construction of a contingency table where a $\chi^2$ test statistic is subsequently calculated. With the use of a Monte Carlo simulation, an estimated {\it p}-value is then obtained for the test statistic. Similar to the model exploration strategy in MDR, the process of contingency table construction and {\it p}-value calculation can also be cycled through all or some of possible interaction models. The model with appropriate candidate SNPs taken from possible SNPs is the one with minimum {\it p}-value and is said to be the best model for interaction explanation. This statistics-based procedure can be found as an integral part of the FAMHAP program~\cite{FAMHAP}.
%
% Fig. 2. VarHAP table
%
\begin{figure}[t!]
\centering
\includegraphics[width=5.0cm]{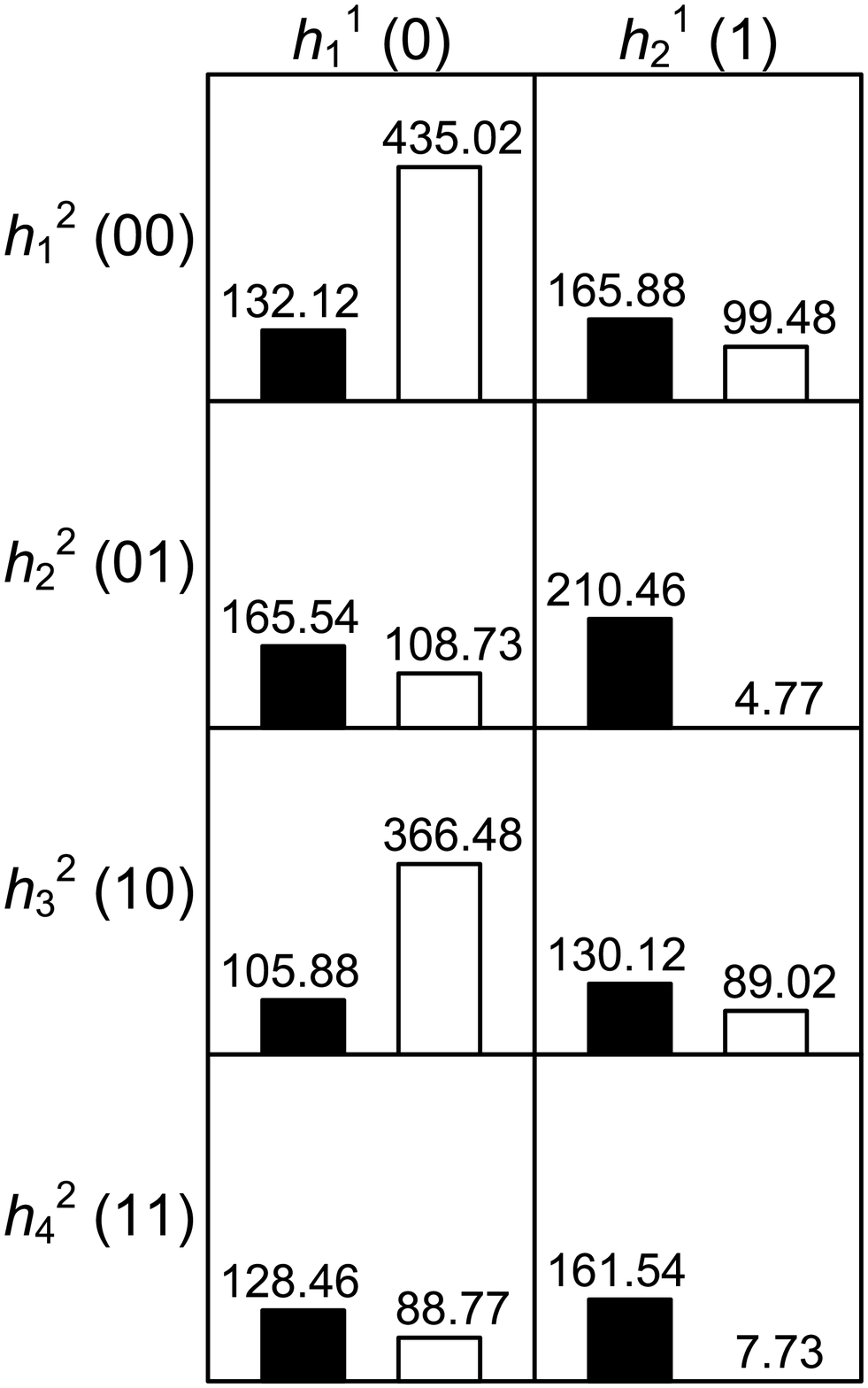}
\caption{A VarHAP decision table which is constructed from 1,200 case-control samples. Haplotypes in the first gene are obtained from one SNP while haplotypes in the second gene are inferred from two SNPs. The table consists of eight cells where each cell represents a unique haplotype configuration. The left (black) bar in each cell represents the accumulative contribution from case samples while the right (white) bar represents the accumulative contribution from control samples. The cells with haplotype configurations $(h_2^1,h_1^2)$, $(h_1^1,h_2^2)$, $(h_2^1,h_2^2)$, $(h_2^1,h_3^2)$, $(h_1^1,h_4^2)$ and $(h_2^1,h_4^2)$ are labelled as predisposing haplotype configurations while the cells with haplotype configurations $(h_1^1,h_1^2)$ and $(h_1^1,h_3^2)$ are labelled as protective haplotype configurations.}
\label{fig:VarHAPTable}
\end{figure}

%
% 3. VarHAP
%
\section{VarHAP}
\label{sec:VarHAP}
VarHAP is proposed for case-control interaction studies. Similar to MDR, the technique is also a classifier-based technique. However, instead of using a genotype data analysis as a means to identify the best SNP combination, the decision table for classification is constructed from the haplotype contribution value described earlier. As a result, haplotypes with different lengths must be inferred during the search for the best model. The number of decision cells during the consideration on haplotypes constructed from a specific set of SNPs is governed by the total number of possible haplotype configurations as illustrated in Figure~\ref{fig:VarHAPTable}. In brief, VarHAP would maintain the ability to find the best SNP combination while also be able to identify possible disease-predisposing and protective haplotype configurations.

Since VarHAP is essentially a classification system, the principal criterion for choosing the optimal SNP combination model is still the prediction accuracy. However, with the use of haplotype contribution value as a means for decision rule construction, an additional model selection criterion that exploits the nature of haplotype can be formulated. This criterion can be referred to as haplotype propagation capability. Basically, if a haplotype constructed from a specific set of SNPs is related to disease susceptibility status, haplotypes constructed from a SNP set which is a superset of the previously specified SNPs should also predict the same relationship. This implies that predisposing and protective haplotypes in a low-order model must be able to propagate into haplotypes in high-order models. For example, consider a single-gene problem with four possible SNPs: \texttt{X1}, \texttt{X2}, \texttt{X3} and \texttt{X4}. If haplotypes in the model with SNPs \texttt{(X2, X4)} are related to disease susceptibility, haplotypes in the models with SNPs \texttt{(X1, X2, X4)}, \texttt{(X2, X3, X4)} and \texttt{(X1, X2, X3, X4)} should produce the same result. The haplotype propagation capability, which is a dichotomous criterion, can be determined from the evidence that the sign test {\it p}-value and the prediction accuracy can be maintained throughout the process of model order increment. Again, in the situation where two or more models with different number of SNPs are equally good in terms of both prediction accuracy and haplotype propagation capability, the most parsimonious model will be the best model.
%
% Table 1. Two-locus disease models
%
\begin{table*}[t!]
\centering
\caption{Description of two-locus disease models. $d_{ij}$ is the penetrance of a genotype carrying $i$ disease alleles at locus 1 and $j$ disease alleles at locus 2. $p_1$ is the frequency of the disease allele at locus 1 while $p_2$ is the frequency of the disease allele at locus 2. $\psi = 2\phi-\phi^2$.}
\begin{tabular}{lcccccccccccc}
\hline
\noalign{\smallskip}
Model & $d_{22}$ & $d_{21}$ & $d_{20}$ & $d_{12}$ & $d_{11}$ & $d_{10}$ & $d_{02}$ & $d_{01}$ & $d_{00}$ & $p_1$ & $p_2$ & $\phi$ \\
\noalign{\smallskip}
\hline
\noalign{\smallskip}
Ep-1 & $\phi$ & $\phi$ & 0 & $\phi$ & $\phi$ & 0 & 0 & 0 & 0 & 0.210 & 0.210 & 0.707 \\
Ep-2 & $\phi$ & $\phi$ & 0 & 0 & 0 & 0 & 0 & 0 & 0 & 0.600 & 0.199 & 0.778 \\
Ep-3 & $\phi$ & 0 & 0 & 0 & 0 & 0 & 0 & 0 & 0 & 0.577 & 0.577 & 0.900 \\
Ep-4 & $\phi$ & $\phi$ & 0 & $\phi$ & 0 & 0 & $\phi$ & 0 & 0 & 0.372 & 0.243 & 0.911 \\
Ep-5 & $\phi$ & $\phi$ & 0 & $\phi$ & 0 & 0 & 0 & 0 & 0 & 0.349 & 0.349 & 0.799 \\
Ep-6 & 0 & $\phi$ & $\phi$ & $\phi$ & 0 & 0 & $\phi$ & 0 & 0 & 0.190 & 0.190 & 1.000 \\
Het-1 & $\psi$ & $\psi$ & $\phi$ & $\psi$ & $\psi$ & $\phi$ & $\phi$ & $\phi$ & 0 & 0.053 & 0.053 & 0.495 \\
Het-2 & $\psi$ & $\psi$ & $\phi$ & $\phi$ & $\phi$ & 0 & $\phi$ & $\phi$ & 0 & 0.279 & 0.040 & 0.660 \\
Het-3 & $\psi$ & $\phi$ & $\phi$ & $\phi$ & 0 & 0 & $\phi$ & 0 & 0 & 0.194 & 0.194 & 1.000 \\
S-1 & $\phi$ & $\phi$ & $\phi$ & $\phi$ & $\phi$ & $\phi$ & $\phi$ & $\phi$ & 0 & 0.052 & 0.052 & 0.522 \\
S-2 & 1 & 1 & 1 & $\phi$ & $\phi$ & 0 & $\phi$ & $\phi$ & 0 & 0.228 & 0.045 & 0.574 \\
S-3 & 1 & 1 & $\phi$ & 1 & $\phi$ & 0 & $\phi$ & 0 & 0 & 0.194 & 0.194 & 0.512 \\				
\noalign{\smallskip}
\hline
\end{tabular}
\label{tab:2-locus}
\end{table*}

%
% 4. Data Sets
%
\section{Data Sets}
\label{sec:DataSets}
The performance of the proposed VarHAP technique is evaluated through benchmark trials. 12 simulated data sets, which represent various gene-gene interaction phenomena including epistasis and heterogeneity, are considered~\cite{Becker,Knapp}. Each data set contains 600 case samples and 600 control samples. Each sample consists of 10 total SNPs from two genes where five SNPs exist in each gene. All SNPs in control samples are in Hardy-Weinberg equilibrium~\cite{Hardy}. Only one SNP from each gene is interacted with one another. The two-locus interaction models are illustrated in Table~\ref{tab:2-locus}. The epistatic models Ep-1--Ep-6 and the heterogeneity models Het-1--Het-3 have been discussed by Neuman and Rice~\cite{Neuman}, who also provide examples of diseases for which these models may be applicable. The heterogeneity models S-1 and S-2 and the epistatic model S-3 have been investigated by Schork et al.~\cite{Schork}. From Table~\ref{tab:2-locus}, if the frequency of the disease allele at a locus is greater than 0.5, the major allele is the disease allele. Otherwise, the minor allele is the disease allele. These interaction models describe disease susceptibility status in terms of penetrance. Penetrance of a genotype with a specific number of disease alleles is the probability that a subject with this genotype has the disease. The test data sets are simulated by a genomeSIM package~\cite{Dudek} with the default setting. As a result, it is also possible to vary the LD pattern among SNPs in the same gene. This leads to two main case studies that need to be explored: strong LD and weak LD cases. In the strong LD case, the susceptibility-causative SNP in each gene and its two adjacent SNPs are in linkage disequilibrium where Lewontin's $D'$ value~\cite{Lewontin} is in the range of 0.80--0.95. In contrast, the Lewontin's $D'$ value for each pairwise LD measurement between susceptibility-causative SNP and its adjacent SNPs is in the range of 0.50--0.60 in the weak LD case. In the strong LD case, an interaction detection technique should be able to identify both the actual two-locus model that directly leads to disease susceptibility and other alternative models which consist of SNPs in strong LD patterns. The ability to detect these other models is important. This is because it is not always straightforward to identify SNPs which are responsible for disease susceptibility in real case-control interaction studies. In contrast, an interaction detection technique should narrow the search to the original two-locus model in weak LD case since it is the only usable model.
%
% Table 2. Weak LD results
%
\begin{table*}[t!]
\centering
\caption{MDR, VarHAP and FAMHAP results from the weak LD case study. 10-fold cross-validation is used in MDR and VarHAP. The prediction accuracy is obtained for the identified principal interaction model. Estimated {\it p}-values in FAMHAP results are equal to zero while sign test {\it p}-values in MDR and VarHAP results are less than 0.001 in all two-locus problems. The technique is said to be able to identify the correct gene-gene interaction model if the reported principal model contains both SNPs which are directly participated in the interaction model. Alternative models are models which contain at least two SNPs where each SNP must be either a SNP from the two-locus model or a SNP which is in linkage disequilibrium with one of the SNPs from the model. The number in each bracket denotes the order of the identified model (the number of SNPs in the model).}
\begin{tabular}{lcccc}
\hline
\noalign{\smallskip}
\multicolumn{1}{c}{Two-} & MDR & VarHAP & Correct & Alternative \\
\multicolumn{1}{c}{Locus} & Prediction & Prediction & Model & Model \\
\multicolumn{1}{c}{Model} & Accuracy & Accuracy & Identification & Identification \\
& (\%) & (\%) & Technique & Technique \\ 
\noalign{\smallskip}
\hline
\noalign{\smallskip}
Ep-1 & 98.00 & 73.92 & MDR(2), VarHAP(2), FAMHAP(2) & FAMHAP(2) \\
Ep-2 & 98.58 & 78.39 & MDR(2), VarHAP(4), FAMHAP(2) & FAMHAP(2) \\
Ep-3 & 99.50 & 87.50 & MDR(2), VarHAP(2), FAMHAP(2) & FAMHAP(2) \\
Ep-4 & 99.25 & 78.96 & MDR(2), VarHAP(2), FAMHAP(2) & FAMHAP(2) \\
Ep-5 & 98.42 & 75.19 & MDR(2), VarHAP(3), FAMHAP(2) & FAMHAP(2) \\
Ep-6 & 100.00 & 85.10 & MDR(2), VarHAP(2), FAMHAP(2) & FAMHAP(2) \\
Het-1 & 93.75 & 73.29 & MDR(2), VarHAP(2), FAMHAP(2) \\
Het-2 & 97.33 & 78.40 & MDR(2), VarHAP(2), FAMHAP(2) & FAMHAP(2) \\
Het-3 & 100.00 & 84.40 & MDR(2), VarHAP(2), FAMHAP(2) & FAMHAP(2) \\
S-1 & 94.00 & 72.98 & MDR(2), VarHAP(2), FAMHAP(2) \\
S-2 & 97.58 & 79.81 & MDR(2), VarHAP(2), FAMHAP(2) & FAMHAP(2) \\
S-3 & 96.75 & 79.15 & MDR(2), VarHAP(2), FAMHAP(2) & FAMHAP(2) \\			
\noalign{\smallskip}
\hline
\end{tabular}
\label{tab:WeakLD}
\end{table*}

%
% 5. Results and Discussions
%
\section{Results and Discussions}
\label{sec:Results}
VarHAP is benchmarked against MDR and FAMHAP. Since the test data contains 10 SNPs, all three techniques have to explore $2^{10} - 1 = 1,023$ possible SNP combination models. An initial investigation reveals that with the use of minimum {\it p}-value as the sole model selection criterion, FAMHAP reports a large number of models with the estimated {\it p}-value equals to zero. As a result, haplotype propagation capability is also implemented as an additional model selection criterion. Further, the parsimony criterion is also utilised when there is a tie between multiple models with different number of SNPs. The results from all three techniques in weak and strong LD case studies are summarised in Tables~\ref{tab:WeakLD} and~\ref{tab:StrongLD}, respectively.

The prediction accuracy of MDR is higher than that of VarHAP in both case studies. This is because VarHAP uses contribution values which are obtained from inferred haplotypes instead of inferred diplotypes---pairs of haplotypes that together describe correct phases of given genotypes---to create decision rules. Consider a situation where disease susceptibility can be determined from a single SNP where the predisposing genotype is the homozygous variant. In other words, the disease susceptibility can be described by a recessive genetic model. MDR can easily classify the heterozygous and homozygous wide-type genotypes as protective genotypes. However, VarHAP would only correctly classify both homozygous genotypes since each genotype is made up from two copies of the same haplotype: two major alleles for the homozygous wide-type and two minor alleles for the homozygous variant. VarHAP would partially misclassify samples with heterozygous genotype. This is because VarHAP identifies the major allele as the protective allele and the minor allele as the predisposing allele. In order to increase the prediction accuracy of VarHAP, it may be necessary to construct decision tables from diplotype information instead of haplotype contribution values. Nonetheless, this will also rapidly increase the dimensions of decision tables in VarHAP.

In the weak LD case study, both MDR and VarHAP are able to identify correct sets of SNPs that lead to disease susceptibility. On the other hand, FAMHAP reports both actual and alternative interaction models. This is undesirable since it would not be possible to further explain disease susceptibility from multiple candidate models in the absence of strong linkage disequilibrium among SNPs. In other words, FAMHAP is quite sensitive in this situation. Further analysis reveals that MDR is marginally better than VarHAP in two epistasis problems: Ep-2 and Ep-5. MDR correctly identifies models which contain two SNPs while the models located by VarHAP contain a few extra SNPs. Nonetheless, these two models identified by VarHAP are still useful to susceptibility explanation.

All three techniques are able to locate correct interaction models in the strong LD case study. However, only FAMHAP and VarHAP are capable of identifying alternative models. Since MDR suggests one candidate model for each fixed-number SNP set, it would not be possible for MDR to produce any alternative models. Recall that these alternative models are equally important since SNPs in the principal two-locus interaction model and SNPs from an alternative model are in strong linkage disequilibrium. This implies that disease susceptibility can be explained using either the original interaction model or the alternative models. This disadvantage in MDR can be overcome if the cross-validation consistency criterion can be replaced by other decision criteria. In this case study, FAMHAP is marginally better than VarHAP in terms of alternative model identification in three epistasis and heterogeneity problems: Ep-3, Ep-6 and Het-3. This means that FAMHAP is at its best when SNPs are in strong linkage disequilibrium. Nonetheless, the overall results from both case studies suggest that VarHAP is the best technique. This is concluded from the fact that VarHAP does not report ambiguous results in weak LD case study while is also capable of producing alternative models in strong LD case study. This is crucial because it is impossible to know beforehand whether susceptibility-causative SNPs are in weak or strong linkage disequilibrium with other SNPs in real case-control interaction studies. In other words, a technique that performs satisfactorily in both weak and strong LD cases would have an advantage over a technique that functions well in only one scenario.
%
% Table 3. Strong LD results
%
\begin{table*}[t!]
\centering
\caption{MDR, VarHAP and FAMHAP results from the strong LD case study. The explanation for how the results are obtained and displayed is the same as that given in Table~\ref{tab:WeakLD}.}
\begin{tabular}{lcccc}
\hline
\noalign{\smallskip}
\multicolumn{1}{c}{Two-} & MDR & VarHAP & Correct & Alternative \\
\multicolumn{1}{c}{Locus} & Prediction & Prediction & Model & Model \\
\multicolumn{1}{c}{Model} & Accuracy & Accuracy & Identification & Identification \\
& (\%) & (\%) & Technique & Technique \\ 
\noalign{\smallskip}
\hline
\noalign{\smallskip}
Ep-1 & 98.00 & 73.92 & MDR(2), VarHAP(2), FAMHAP(2) & VarHAP(2), \\
 & & & & FAMHAP(2) \\
Ep-2 & 98.58 & 77.02 & MDR(2), VarHAP(4), FAMHAP(2) & VarHAP(4), \\
 & & & & FAMHAP(2) \\
Ep-3 & 99.50 & 87.50 & MDR(2), VarHAP(2), FAMHAP(2) & FAMHAP(2) \\
Ep-4 & 99.25 & 78.96 & MDR(2), VarHAP(2), FAMHAP(2) & VarHAP(2), \\
 & & & & FAMHAP(2) \\
Ep-5 & 98.42 & 75.87 & MDR(2), VarHAP(3), FAMHAP(2) & VarHAP(3), \\
 & & & & FAMHAP(2) \\
Ep-6 & 100.00 & 85.10 & MDR(2), VarHAP(2), FAMHAP(2) & FAMHAP(2) \\
Het-1 & 93.75 & 75.41 & MDR(2), VarHAP(3), FAMHAP(2) & VarHAP(3), \\
 & & & & FAMHAP(2) \\
Het-2 & 97.33 & 78.40 & MDR(2), VarHAP(2), FAMHAP(2) & VarHAP(2), \\
 & & & & FAMHAP(2) \\
Het-3 & 100.00 & 84.40 & MDR(2), VarHAP(2), FAMHAP(2) & FAMHAP(2) \\
S-1 & 94.00 & 72.98 & MDR(2), VarHAP(2), FAMHAP(2) & VarHAP(2), \\
 & & & & FAMHAP(2) \\
S-2 & 97.58 & 79.81 & MDR(2), VarHAP(2), FAMHAP(2) & VarHAP(2), \\
 & & & & FAMHAP(2) \\
S-3 & 96.75 & 79.15 & MDR(2), VarHAP(2), FAMHAP(2) & VarHAP(2), \\
 & & & & FAMHAP(2) \\			
\noalign{\smallskip}
\hline
\end{tabular}
\label{tab:StrongLD}
\end{table*}

%
% 6. Conclusions
%
\section{Conclusions}
\label{sec:Conclusions}
In this paper, a non-parametric pattern recognition/classification technique for case-control gene-gene interaction studies is presented. Instead of using direct genotype inputs in classification, inferred haplotypes, which are obtained through an expectation-maximisation algorithm~\cite{Excoffier}, are used as inputs. Each case/control sample contributes values derived from inferred haplotypes to decision tables which are constructed and tested for all possible gene-gene interaction models. The technique primarily uses prediction accuracy obtained from {\it k}-fold cross-validation as a means for identifying candidate SNPs which are responsible for disease susceptibility. The technique also employs haplotype propagation capability as an additional criterion. If the selection procedure ends in a tie between two or more models with different number of SNPs, the most parsimonious model is then reported as the interaction model. Since haplotypes with different length must be constructed during model identification, the proposed technique can be referred to as a variable-length haplotype construction for gene-gene interaction (VarHAP) technique. VarHAP has been benchmarked against two interaction model detection programs namely MDR~\cite{Hahn} and FAMHAP~\cite{Becker,FAMHAP} in 12 two-locus epistasis and heterogeneity problems~\cite{Becker,Knapp}. The results reveal that FAMHAP reports multiple ambiguous models in the presence of weak linkage disequilibrium among input SNPs while MDR is not suitable for alternative interaction model identification when input SNPs are in strong linkage disequilibrium. In contrast, VarHAP emerges as the most suitable technique in both situations involving weak and strong linkage disequilibrium. Suggestions for further improvement of MDR and VarHAP are also included.
%
% Supplementary Information
%
\section*{Supplementary Information}
VarHAP, which is implemented in Java, and the simulated data sets used in the article are available upon request \texttt{(email: nchl@kmutnb.ac.th)}. In addition to the use of the genomeSIM package~\cite{Dudek}, the data sets can also be generated by a SNaP package~\cite{Nothnagel}. Readers might also be interested in applying the techniques discussed in this article to examples of case-control data sets, which are publicly available from the Wellcome Trust Case Control Consortium~\cite{WTCCC}.
%
% Acknowledgements
%
\section*{Acknowledgments}
A. Assawamakin was supported by the Thailand Research Fund (TRF) through the Royal Golden Jubilee Ph.D. Programme (Grant No. PHD/4.I.MU.45/C.1) and the National Center for Genetic Engineering and Biotechnology (BIOTEC), the National Science and Technology Development Agency (NSTDA). N. Chaiyaratana was supported by the Thailand Research Fund. C. Limwongse was supported by the Mahidol Research Grant. P. Youngkong was supported by the Commission on Higher Education (CHE). The authors acknowledge S.M. Dudek at the Vanderbilt University for providing an access to the genomeSIM package.
% 
% References

\end{document}